# ANN Model to Predict Stock Prices at Stock Exchange Markets

**Wanjawa, Barack Wamkaya**

School of Computing and Informatics,

University of Nairobi,

wanjawawb@gmail.com

**Muchemi, Lawrence**

School of Computing and Informatics,

University of Nairobi,

lmuchemi@uonbi.ac.ke

Abbreviated title: **ANN Model for Stock Market Prediction**

**Corresponding author:**

**Wanjawa, Barack Wamkaya**

School of Computing and Informatics,

University of Nairobi, Box 30197,

Nairobi 00100 Kenya

wanjawawb@gmail.com

Tel. +254722614787

**ABSTRACT**

Stock exchanges are considered major players in financial sectors of many countries. Most Stockbrokers, who execute stock trade, use technical, fundamental or time series analysis in trying to predict stock prices, so as to advise clients. However, these strategies do not usually guarantee good returns because they guide on trends and not the most likely price. It is therefore necessary to explore improved methods of prediction.

The research proposes the use of Artificial Neural Network that is feedforward multi-layer perceptron with error backpropagation and develops a model of configuration 5:21:21:1 with 80% training data in 130,000 cycles. The research develops a prototype and tests it on 2008-2012 data from stock markets e.g. Nairobi Securities Exchange and New York Stock Exchange, where prediction results show MAPE of between 0.71% and 2.77%. Validation done with Encog and Neuroph realized comparable results. The model is thus capable of prediction on typical stock markets.

## 1.0     INTRODUCTION

Stock markets are trading institutions where stocks (equity) and other financial instruments such as bonds are offered for trade. For stocks, the market generally operates a 'willing-buyer, willing-seller' trade, where buyers and sellers prices are matched for a fit. If there is no match, then no trade takes place and waits for a future match or expires. In most stock exchanges, the common and easily accessible market is the equity market (stocks), where the entry investment can be as low as USD1. The equity market is therefore more active, having many players and hence a segment worthy of further study. The performance of stock markets is measured on a daily basis by some key indicators such as 'share index', which is a measure of the performance of some stocks picked from the different sectors of the market. Such an index is important in not only gauging the performance of trades in the stock exchange but also the economic performance of the particular country as a whole.

Shareholders however do not directly execute the trade, nor is there any meeting between buyers and sellers for negotiations. Shareholders trade by giving instructions to their Stockbrokers, who in turn execute the orders. Stockbrokers usually also advise clients on where to trade. In their advisory role, some Stockbrokers base their advice on the fundamentals of the various stocks or undertake technical analysis. However, none of these predictive methods have assurance of profit as they usually just indicate a future trend and a likely up or down price movement and not the real expected future stock price. Stockbrokers need to be empowered, through better predictive tools, to enable them have some capability to provide the best advice to their clients. A predictive tool that Stockbrokers can use to guide on exact price movements, as a basis of investment, is therefore desirable. This can be an artificial intelligence (AI) system based on neural networks.

Due to the importance of stock markets, investment is usually guided by some form of prediction. However, predicting the stock market is not a trivial task. To start with, there is need to model the trend of the stock prices, which is nonlinear. Additionally, it is desirable to extract certain features on the input data itself, to make it capable of good use (Zarandi et al. 2012). NeuroAI (2013) states that there are four stock market prediction methods. The first is technical analysis. Huang et al. (2011) states that technical analysis deals with historical price movement to predict a pattern for future investment decision. Deng et al. (2011) lists some of the popular technical indicators as rate of change (ROC), Moving Average Convergence Divergence (MACD) and bias. The second is fundamental analysis, which according to Chen et al. (2007) is the investigation of relationship between financial information and other facts about the company such as inventory or revenue growth.

A third method of stock market prediction is Time series method, which uses historical performance to predict on a time series scale. A time series is a sequence of sampled quantities from an observation out of which discoveries such as periodic distribution can be determined (Zhang et al. 2008). Other methods in time series prediction are linear regression, auto-regression and Auto-regression Integrated Moving Average (ARIMA). An important characteristic of time series data is the dependence on time, hence current observations depend on past observation in time (Neto et al. 2009). The last method is the use of Artificial Intelligence (AI). AI problems are usually solved using agents, such as the learning agent. Learning allows an agent to become more competent than its initial knowledge. There are three types of learning, depending on the feedback that the agent receives. In Unsupervised learning, the agent learns of patterns even without explicit feedback, while in Reinforcement learning the agent's knowledge is reinforced by rewards and punishment. Finally, in Supervised learning, the agent is provided with data so that it can observe input and output pairs so as to formulate a function to map such pairs as applied in Genetic Algorithm (GA) and Artificial Neural Networks (ANN).

An ANN is modeled as a representation of the humans own biological neuron and how the neural network functions in a human being. An ANN is an AI computation method that is also called connectionism, parallel distributed processing or neural computation. One reason why ANN is popular is due to its robustness, fault tolerance, ability to learn and generalize, adaptability, universal function approximation and parallel data processing. This enables them to solve complex non-linear and multi input-output (IO) relationship problems (Ortiz-Rodriguez et al., 2013 and Gomes et al., 2011) Ghaffari et al. (2006) states that when compared to other methods, ANN have been shown to be superior as modeling technique for data sets with non-linear relationships for applications such as data fitting and prediction. Cerna et al. (2005) states that the most popular types of ANN are the Kohonen network (self-organizing map) and the multi-layer perceptron (MLP), which use unsupervised and supervised learning respectively. Ortiz-Rodriguez et al. (2013) states that multilayer perceptron trained with backpropagation algorithm is the 'most used ANN' in prediction, classification and modeling. Training can be fully supervised learning as used in classification or semi-supervised learning as used in clustering. There is usually need to measure system performance in typical prediction applications. In other research, Deng et al. (2011) measured performance of their model on selected stocks at the New York Stock Exchange on the basis of Mean Absolute Percentage Error (MAPE) and Root Mean Square Error (RMSE), while Neto et al. (2009) also measured performance of their model for a Brazilian stock company using MAPE and Mean Square Error (MSE).

A challenge in ANN design is the selection of the optimum number of units that are large enough to fit the purpose but not too large that the ANN fails to generalize the solution (over-fitting). There are several issues to consider in the design of ANNs, such as: type of network (feedforward, recurrent, backpropagation), type of training (supervised, unsupervised), proportion of training and testing data sets (70:30 or 80:20), number of input and output units (usually application dependent), number and size of hidden layers (2N+1, experimental), number of repetitions during training (epoch), choice of activation function (sigmoid, linear, hyperbolic, threshold) and size of data set (number of records). Other design considerations include data volume, learning rate and momentum. In determining appropriate data volume, Devi et al. (2011) did tests over a three year period while Adhikari et al. (2013) used datasets of between 3 years and 4 years. On their part, Wong et al. (2012) considered a four year period while Butler et al. (2011) used data that was upto 23-years. For training and testing data ratios, Ortiz-Rodrigues et al. (2013) used a 80:20 ratio, while Microsoft (2013) uses 70:30 on their SQL server product.

Research has been done on other markets, where stock market price prediction has been attempted. Deng et al. (2011) applied technical analysis for prediction of select stocks at the New York Stock Exchange. Aghababaeyan et al. (2011) undertook study on prediction for the Tehran Stock Exchange, where they developed a tool that achieving an accuracy of 97%. Others such as Khan et al. (2011) conducted a study on the Bangladesh Stock market where they developed a tool with average error of 3.7% & 1.5% in two simulations, while Pan et al. (2005) developed a tool for the Australian Stock market with 80% accuracy in predicting the price direction. None of their tools seem to have been developed commercially or targeted for the respective stockbrokers.

**Problem Statement:** Trading in shares is big business in many economies. Based on the information on their websites, Stockbrokers do not seem to have any intelligent tool that can help them advise clients on which stocks are suitable for any buy or sale trade. These websites provide information that points to use of fundamental, technical and time series analysis methods, as depicted in Figure 1.1 below. These prevalent methods show a trend on future movement and not the likely trade price for any stock in future. It is therefore desirable to have a tool that does not just point a direction of price movement, but also indicates the most likely price value of the stock itself. An ANN model that is well tuned with the appropriate parameters can be used to develop such a predictive tool.

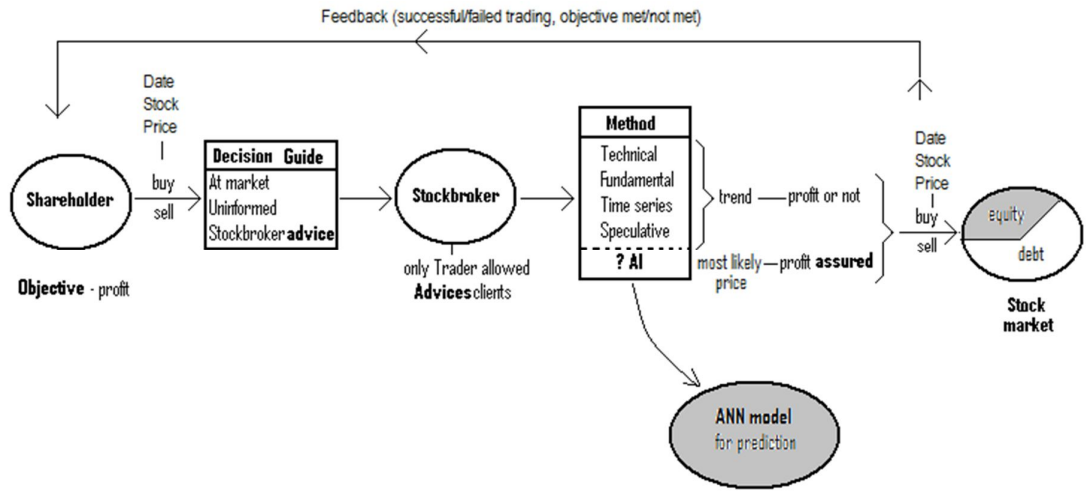

*Figure 1.1* – *Illustration of the Stock Trade Cycle and Methods used*

## 2.0    ANN-BASED STOCK PREDICTION MODEL

### 2.1    Baseline ANN Stock Prediction Model

This research designed a tuned AI model, based on the ANN algorithm and developed the model into a working prototype. The specific AI model agent was a feedforward ANN with multilayer perceptron using backpropagation and trained using supervised learning. The programming language environment was C# (C# Express 2010 version).

The baseline ANN model was subjected to both a training phase and a testing phase from the available data. The data used to test the tool was the daily closing price of individual stocks (about 60 companies) traded at the Nairobi Stock Exchange (NSE) over a five year period, 2008 to 2012 (1,000 data rows), obtained from secondary sources (Synergy 2013). The initial baseline model used 70:30 training and testing data ratios, with a configuration of 5:11:11:1, as guided by the formulae 2N+1 as a best practice. 70% training data was the period from January 2, 2008 to July1, 2011. The project tested predictions over a 3-month period (the next 60 predictions) from the remaining pool of 30% data.

On a 5:11:11:1 configuration, the number of inputs are 5 i.e. 5 daily stock prices with the aim of predicting the 6$^{th}$ price in the series. Other settings for the baseline model were: Number of hidden layers = 2, Number of neurons per hidden layer = 11, Number of outputs = 1, Bias per layer = 1. The baseline design is shown on **Figure 2.1** below. Pseudocode and flowcharts were used to assist in developing the C# program. The final product was a C# program with an interface that allows for a Training action (Train tab) and a Testing action (Predict tab).

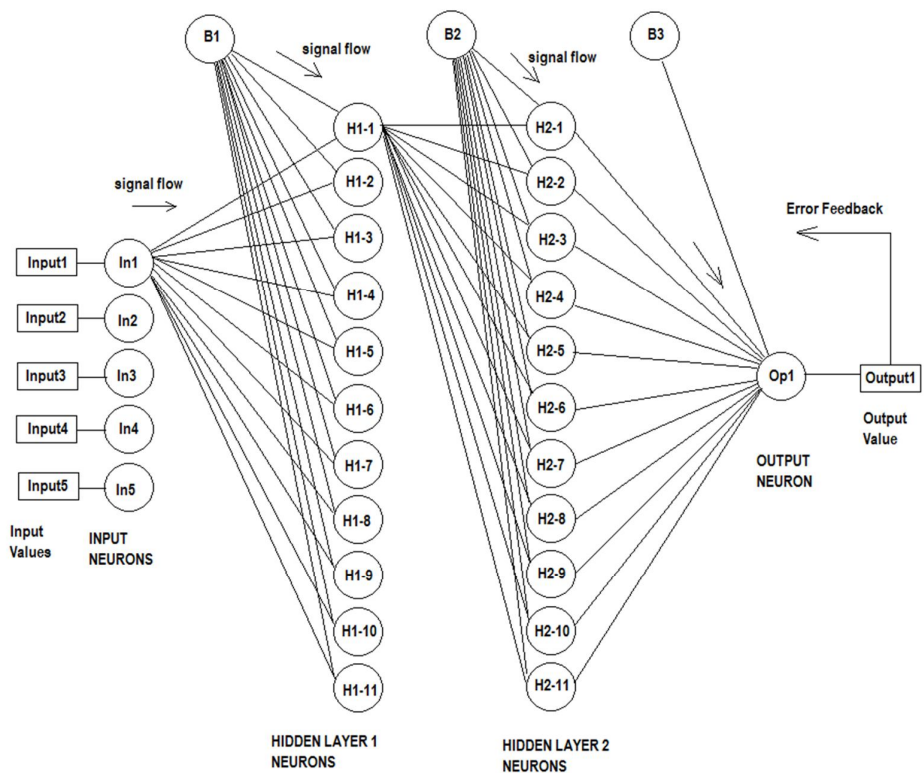

For clarity, neuron to neuron connection is only shown for In1, H1-1, H2-1, Op1 and biases B1, B2, B3. However, each neuron connects to others in same scheme e.g. In2 connects to H1-1 to H1-11 etc., while H1-2 connects to H2-1, H2-2 etc.

*Figure 2.1 – The Baseline ANN design Model*

Since it was not possible to test all the sixty listed stocks at the NSE, stocks for this research were chosen from those that met some criteria. The considerations were: the stock was part of the 20-share index (barometer of the economic), it was consistently traded in the 5-year period and the stock had not been subjected to a share split or bonus share issue in the period. Six stocks met these requirements: Stock01 (Kakuzi), Stock02 (Standard Bank), Stock03 (Kenya Airways), Stock04 (Bamburi), Stock05 (Kengen) and Stock06 (BAT).

The baseline test resulted into the varying MAPE per test stock. This is shown on the graph in **Figure 2.2** below. The baseline test was used to determine one of the six stocks, with the best mean absolute percentage error (MAPE), which was then used to conduct the other tuning and data volume determination tests that were used to develop the new model. This one stock used for other tune-up experiments was Stock02 (Standard Bank).

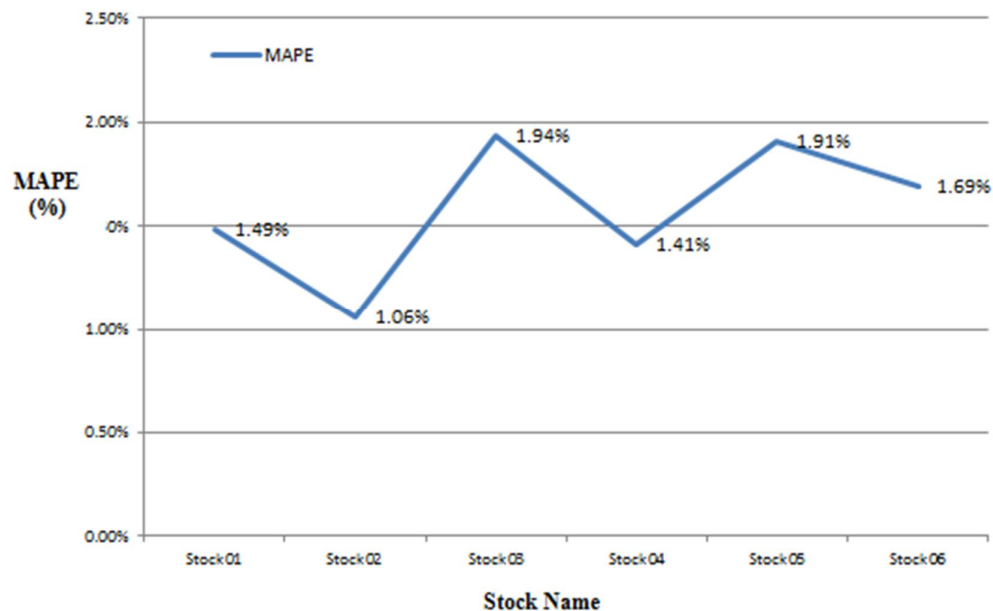

*Figure 2.2* – *MAPE obtained when testing the 6 stocks on Baseline ANN settings*

## 3.0 TUNING PARAMETERS OF THE BASELINE MODEL

The process of tuning the parameters of the baseline model is shown on the flowchart in **Figure 3.1** below.

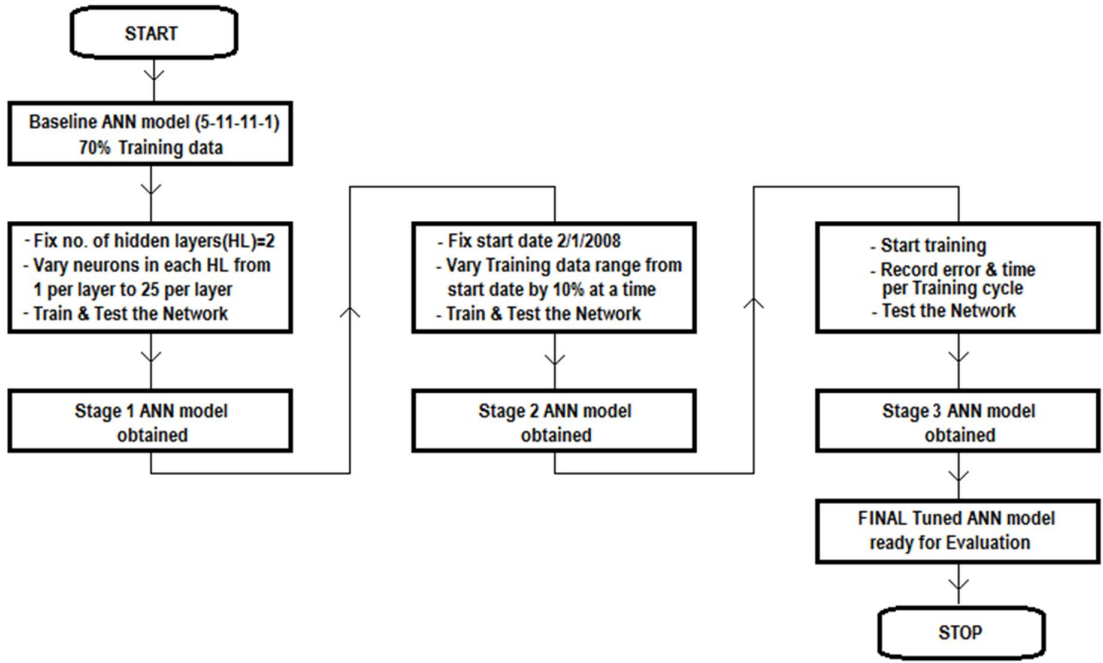

*Figure 3.1* – *Process of Tuning the ANN Baseline Model to Obtain a Final Model*

### 3.1 The Tuning Process

The baseline model of settings 5:11:11:1 was tuned, through experiment, to determine a new optimum model. The data volume available was from NSE for 5-year period January 2008 to December 2012. The prediction period was fixed at 3 months or the next 60 predictions after end of training date. The experimental evaluation used Stock02, which achieved the lowest MAPE based on the design time baseline. Several experimental stages were carried out. The first experiment was done by setting the number of hidden layers fixed at two and then progressively adjusting the number of neurons per hidden layer, from 1 neuron per layer to 25 neurons per layer. After each adjustment, a series of training and testing phases were done.

The second experiment was done based on the model from first experiment. The start date was fixed as Jan. 2, 2008 and then 10% of the available data was used for training, followed by testing of the network on predicting the next 60 values. Training data volume was then progressively increased from 10% to 90% in increments of 10%. The final experiment was done on the network obtained from the second experiment. This final experiment tested the error rate obtained per training cycle, from 0 to 180,000 training cycles.

## 3.2    Results of the Tuning Experiments

The results obtained for the first experiment of varying the number of neurons per hidden layer, are shown graphically on **Figure 3.2**.

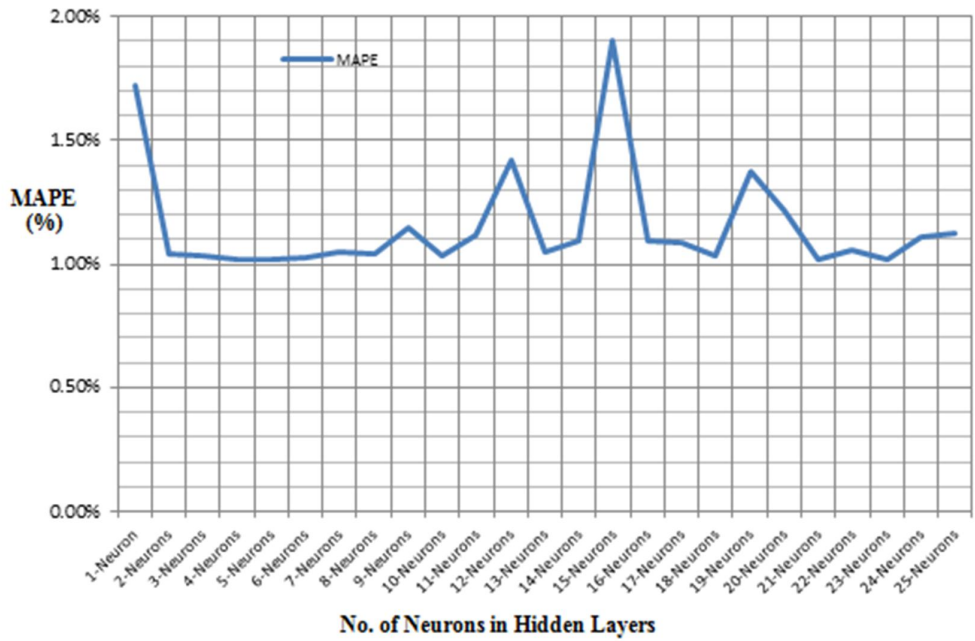

*Figure 3.2 – Stock02 (Standard Bank) MAPE for different settings of neurons per each of the 2 hidden layers*

The results obtained for the second experiment of varying the training data volume from 10% to 90% are shown graphically on **Figure 3.3** below.

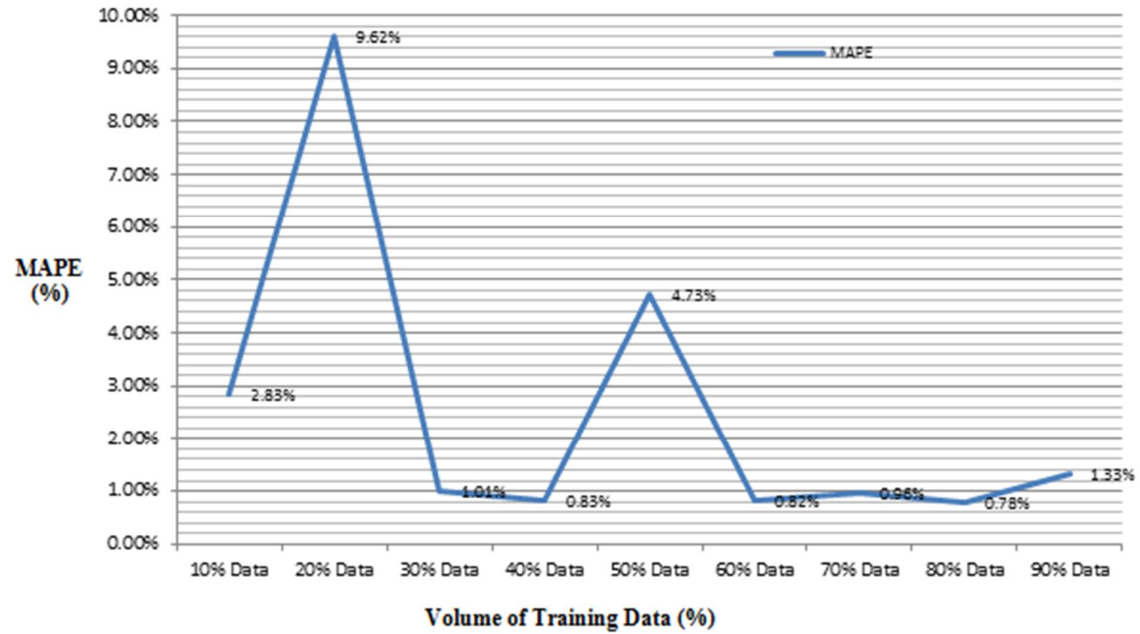

*Figure 3.3* – *MAPE for different volumes of Training data*

The results obtained for the final experiment of varying the number of training cycles from 0 to 180,000 repetitions are shown on the graph on **Figure 3.4** below.

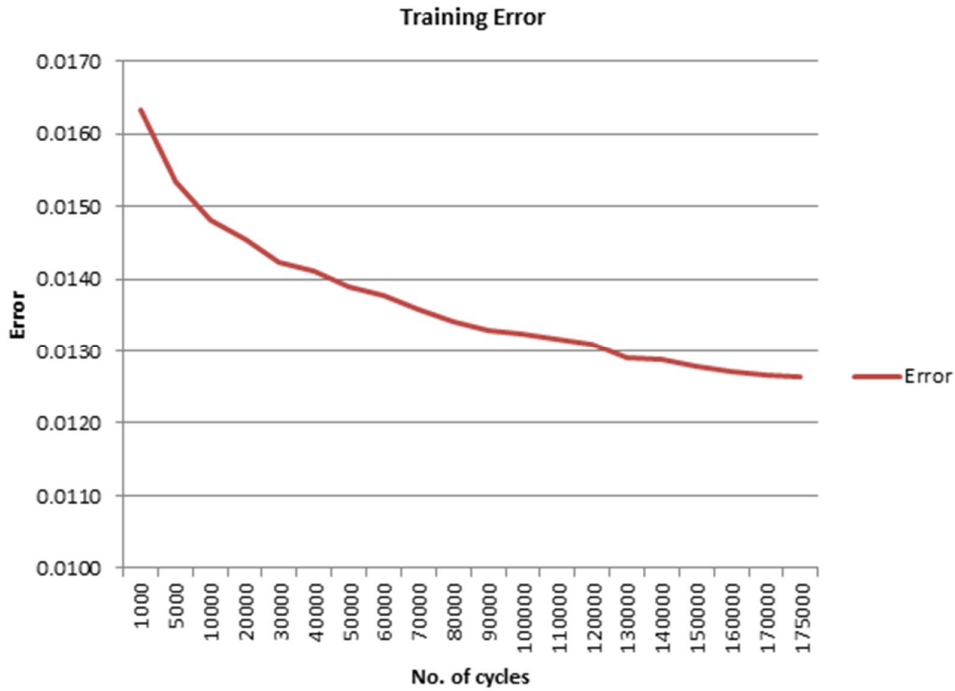

*Figure 3.4 – Training error for different cycles of Training data*

### 3.3   The Final Tuned ANN-Model

The final model, as determine by experimentation was of configuration of 5:21:21:1, using 80% of the available data (Jan. 2, 2008 to Dec. 31, 2011) for training, with at least 130,000 repetitions during training (about 1hour 40min). The testing period was 3 months, from January 2, 2012 to Mar. 31, 2012. The new developed model is shown in **Figure 3.5** below.

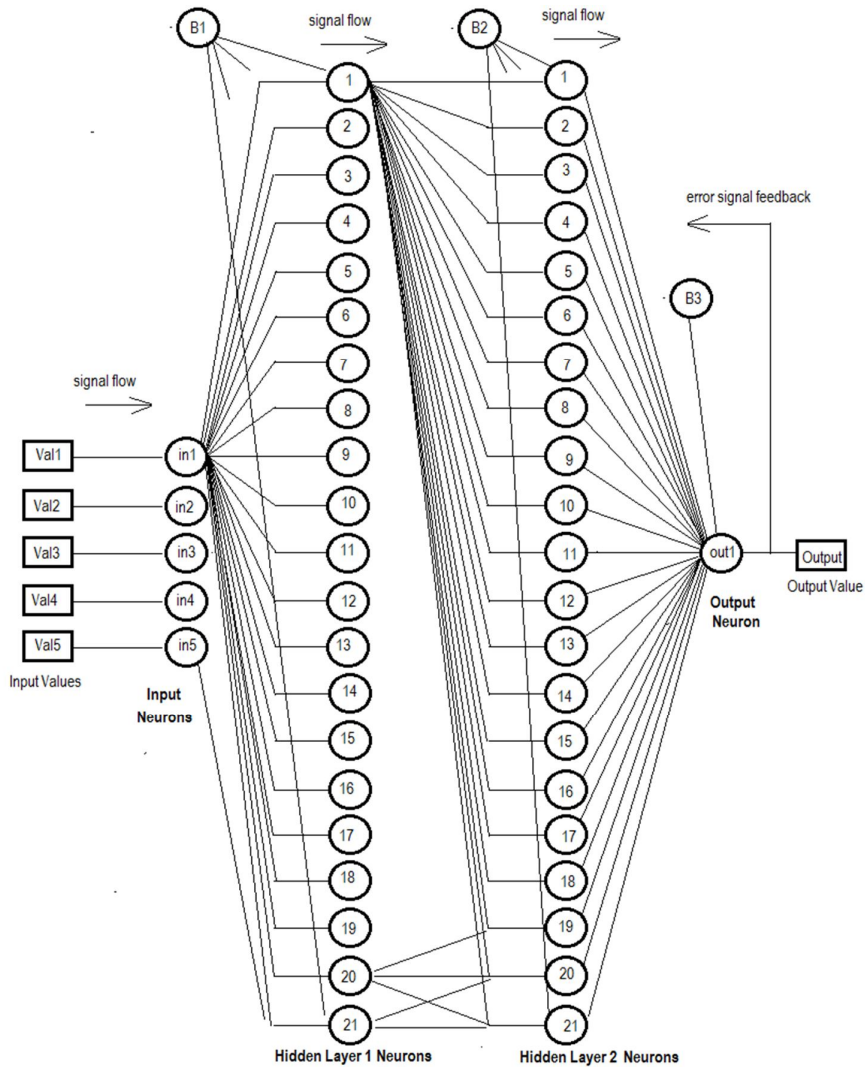

For clarity, only typical neuron to neuron connections is shown. In the model, each neuron connects to the next layer e.g. Input Layer neuron1 connects to each Hidden Layer neuron, same to Input neuron 2 etc. Hidden Layer 1 neuron 1 connects to Hidden Layer 2 neuron 1 to 21, same to neuron 2 etc.

*Figure 3.5* – *The New ANN model For Stock Market Prediction*

## 4.0 EVALUATION OF THE MODEL

### 4.1 Evaluation Process

Three tests were done to evaluate the new model. The first test was to create a prototype based on the new model, using C#. This prototype was used to test some stocks from the Nairobi Stock Exchange. Three out of the six shortlisted stocks were used in this evaluation i.e. Stock01 (Kakuzi), Stock02 (Standard Bank) and Stock04 (Bamburi Cement). The second evaluation was to conduct comparative tests using two other open source tools i.e. Neuroph version 2.8 (201306052037) & Encog Workbench version 3.0.1, with Java Runtime Environment (JRE) version 1.7.0_45, based on the same three NSE stocks on the same time period, using the same network configuration as the model. Root mean squared error (RMSE) was used for comparing the performance of the three different tools on the same data domain. For the final evaluation, the prototype was used to test some stocks traded at the New York stock exchange, whose data was available in the public domain (Yahoo Financials, 2013). Three popular stocks were selected for this test in the same time period 2008 to 2012. These were: Stock01 (Microsoft), Stock02 (Coca-Cola) and Stock03 (Alcatel). It was assumed that the issue of splits, bonuses and continuity had been addressed in the datasets as obtained.

## 4.2    Results and Discussions

The first test was done by using the tuned model prototype on NSE data (2008 to 2012). The MAPE results obtained using the prototype based on the developed model (5:21:21:1, with 80% for testing) is shown graphically in **Figure 4.1** below. It was noted that the error (MAPE) when using the prototype to predict the actuals was 0.77%. The prediction trend was also seen to follow that of the actual trades. The highest variance on any day of trade in the 64-day period was 4.3% (22-Mar-2012)

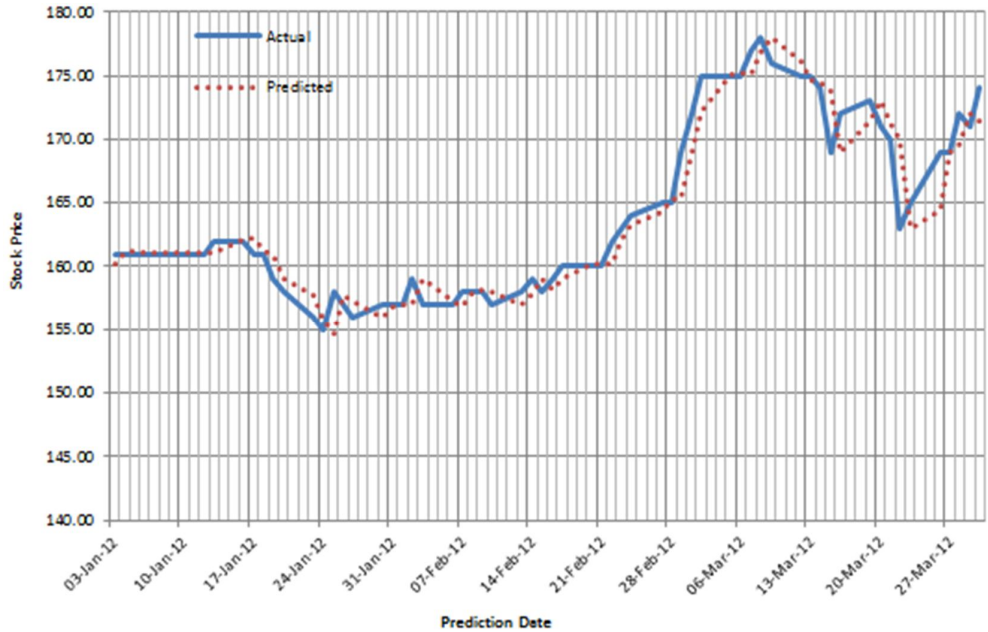

*Figure 4.1* – *Stock02 Results (Using ANN Prototype) on NSE data – Comparing Actual Against Predicted*

The second test was comparative evaluation of the new model prototype against two other open source tools on the same NSE data. MAPE results obtained using Encog Workbench and Neuroph as compared to the prototype for one of the stocks (Stock02-Standard Bank) is shown graphically in **Figure 4.2** below. It was observed that for all the three test tools, the prediction trend was that of the actual trades for all observations. With RMSE of 1.83, the prototype was the most accurate of the three tools. Encog Workbench and Neuroph achieved an RMSE of 1.94 and 2.85 respectively.

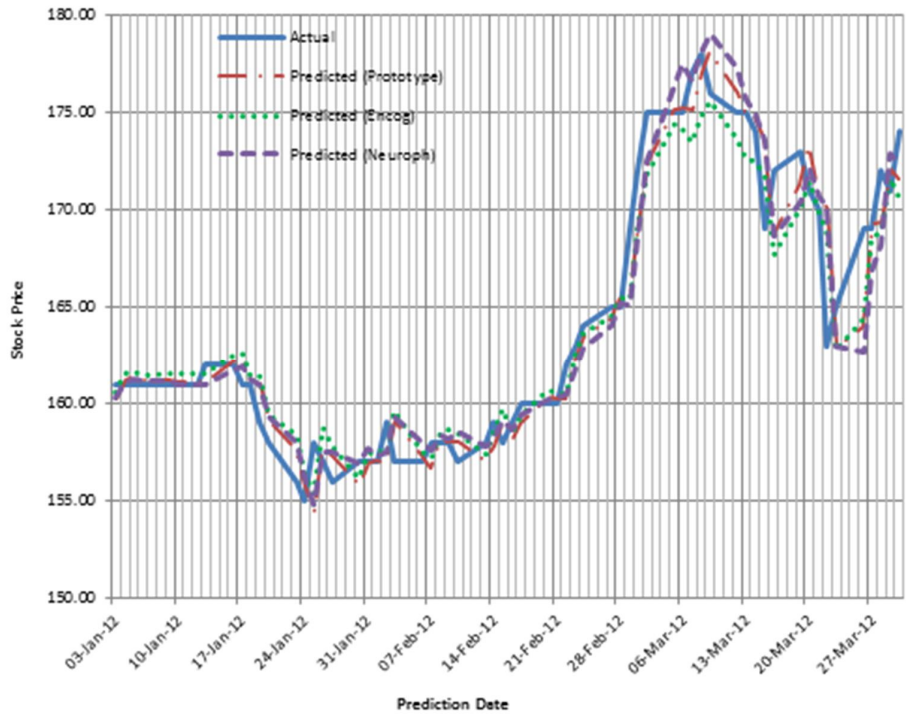

*Figure 4.2* – *Stock02 Results (Comparison of the 3 tools) on NSE data – Comparing Actual Against Predicted*

The final test results obtained using the ANN prototype on data from New York Stock Exchange, for one of the stocks (Stock02 Coca Cola) is shown graphically in **Figure 4.3** below. It was observed that the error (MAPE) when using the prototype to predict the actual was 0.71%. The highest variance on any day of trade in the 62-day period was -3.2% (29-Mar-2012)

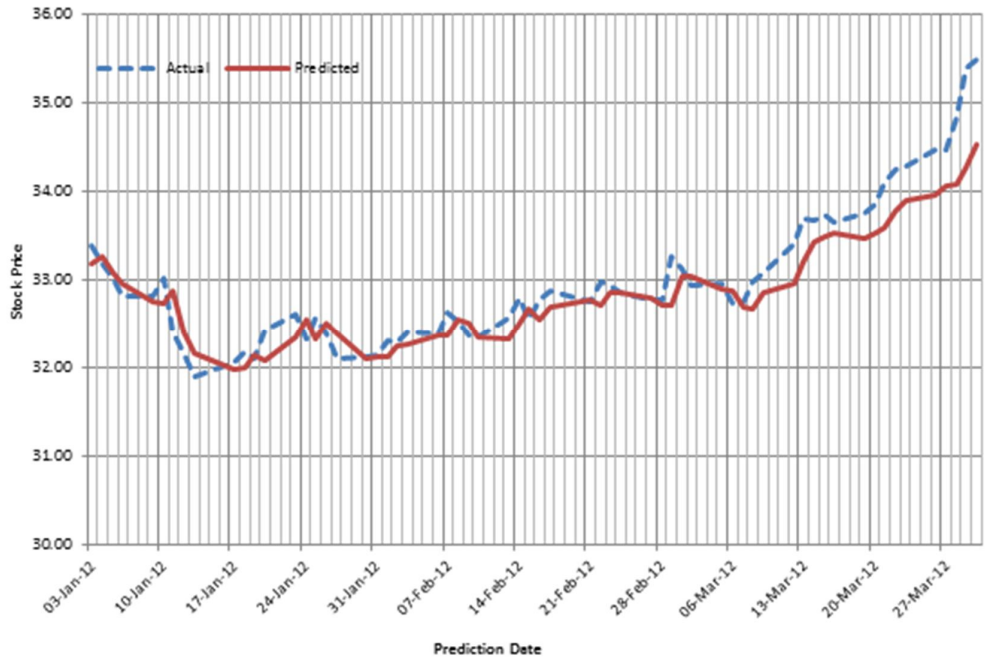

*Figure 4.3* – *Stock02 (Coca Cola) Results (Using Prototype) – Comparing Actual Against Predicted*

The results obtained indicated that it was possible to develop an ANN model that could be used for the prediction of stocks at a typical stock market. This ANN model was a multi-layer perceptron (MLP) using a feedforward network, trained using error backpropagation. The design of the model started with a baseline configuration of 5:11:11:1, using 70% of the available data for training. The baseline design was then further tuned experimentally by first determining an optimum ANN architecture, which resulted into a 5:21:21:1 network, as stage 1. Using this optimized network, from stage 1, the research then tested the effect of varying the training data volume on the prediction accuracy at testing stage 2. The best performance was achieved with 80% data volume used for training, indicating that at least 1,000 training records were necessary for predicting the next 60 days. The model was also tested through a third stage, where it was determined that at least 130,000 training cycles were necessary to achieve acceptably low training error. This took about 1hr 40min of training.

When the prototype was tested on three selected NSE stocks, where it achieved error (MAPE) of 0.77%, 1.44% and 1.91% in the order of error values for Stock02 (Standard Bank), Stock04 (Bamburi Cement) and Stock01 (Kakuzi) respectively. From comparative analysis with other existing tools, the developed ANN prototype was capable of desirable performance. Based on RMSE comparison of the three tools (Prototype, Neuroph and Encog), the prototype performed almost at the same level as the other two tools on the three chosen stocks. The respective RMSE for the Prototype, Encog and Neuroph were 2.26, 2.19 & 2.18 for Stock01; 1.83, 1.94 and 2.85 for Stock02 and 3.07, 2.81 and 3.04 for Stock04. These were quite close, with an extreme difference of only 9%.

The prototype was tested for adaptability to other stock markets using data from the New York Stock Exchange, for data of the same period, for three selected stocks. It was observed that the predicted stock price values followed the trend of the actual prices on the respective dates. Based on the results, where the MAPE was between 0.71% and 2.77%, the model was found capable of use on such other stock exchanges.

## 5.0    CONCLUSION

It is possible to use Artificial intelligence (AI) to develop models that can be used in prediction. Such models are applicable to the financial markets such as the stock exchange. The AI method that was found suitable for these models was the Artificial Neural Network (ANN), which exploits parallel computing to gain intelligence from input data as a basis of predicting future values. These ANN models, such as the multi-layer perceptron (MLP) using feedforward network with error backpropagation, can be developed into prototypes using typical programming languages such as C#. ANN tools need substantial data for training to enable them have the capacity of prediction. The research determined that a model of configuration of 5:21:21:1 achieved the highest prediction accuracy. The research also determined that at least 1,000 records, which was 80% of the data, training over 130,000 cycles, was needed for the training set to achieve best results for a prediction of 60 future values (3-months). Once trained, the ANN-based system was capable of very high precision in its prediction. Validation of the model was done using two open source tools (Encog Workbench and Neuroph).

Based on the 2008-2012 data, the project prototype, based on the above model, achieved a Mean Absolute Percentage Error (MAPE) that were as low as 0.79% on one of the three test stock, Stock02 (Standard Bank). The highest MAPE was 1.91% on Stock01 (Kakuzi). The developed prototype also compared quite favorably with the other open source tools that were identified for comparative analysis and validation i.e. Neuroph and Encog Workbench. Based on MAPE, all tools performed almost at par. The prototype was observed to be generic enough to find applicability not only to the NSE but to other stock exchanges as well. Sample test done on three selected stocks from the New York stock exchange (NYSE) showed that the model was capable of good prediction based on the same period (2008-2012) with error (MAPE) ranging from 0.71% to 2.77%.

For further studies, there is need to formulate a model that is generic enough to suit the full range of stocks at the stock market. This research carefully selected six stocks as the basis of the study, though the total listed companies were sixty. There is also need to explore ANN configurations above the 25 neurons studied in this research and also the effect of increasing the number of hidden layers from the two that were studied in this research. Finally, further research is needed to determine how long a trained ANN system remains valid and effective in prediction before it is found to be in need of retraining.

**REFERENCES**

1. Adhikari R., and R. K. Agrawal. 2013. A Combination of Artificial Neural Network and Random Walk Models for Financial Time Series Forecasting. *Neural Computing and Applications*

2. Aghababaeyan R., and N. TamannaSiddiqui. 2011. Forecasting the Tehran Stock Market by Artificial Neural Network. *International Journal of Advanced Computer Science and Applications, Special Issue on Artificial Intelligence*

3. Butler M., and A. Daniyal. 2009. Multi-objective Optimization with an Evolutionary Artificial Neural Network for Financial Forecasting. *Proceedings of the 11th Annual conference on Genetic and evolutionary computation* 1451-1457

4. Cerna L., and M. Chytry. 2005. Supervised Classification of Plant Communities with Artificial Neural Networks. *Journal of Vegetation Science* 16:407-414

5. Chen Y., and C. Cheng. 2007. Forecasting Revenue Growth Rate Using Fundamental Analysis: A Feature Selection Based Rough Sets Approach. *Fourth International Conference on Fuzzy Systems and Knowledge Discovery (FSKD 2007)* 3:151-155

6. Deng, S., T. Mitsubuchi, K. Shioda, T. Shimada and A. Sakurai. 2011. Combining Technical Analysis with Sentiment Analysis for Stock Price Prediction. *2011 IEEE Ninth International Conference on Dependable, Autonomic and Secure Computing* 800-807

7. Devi, B. U., D. Sundar and P. Alli. 2011. A Study on Stock Market Analysis for Stock Selection – Naïve Investors' Perspective using Data Mining Technique. *International Journal of Computer Applications* 34(3):19-25

8. Ghaffari, A., H. Abdollahi, M. R. Khoshayand, I. S. Bozchalooi, A. Dadgar and M. Refiee-Tehrani. 2006. Performance Comparison of Neural Network Training Algorithms in Modeling of Bimodal Drug Delivery. *International Journal of Pharmaceutics* 327(1-2):126-138

9. Gomes, G. S., T. B. Ludermir, and L. M. Lima. 2011. Comparison of New Activation Functions in Neural Network for Forecasting Financial Time Series. *Neural Computing and Applications* 20(3):417-439

10. Huang, C., P. Chen and W. Pan. 2011. Using Multi-Stage Data Mining Technique to Build Forecast Model for Taiwan Stocks. *Neural Computing and Applications* 21(8):2057-2063.

11. Khan, Z. H., T. S. Alin and A. Hussain. 2011. Price Prediction of Share Market using Artificial Neural Network (ANN). *International Journal of Computer Applications (0975–8887)* 22(2)

12. Microsoft. 2013. MSDN – Training and Testing Data Sets. http://msdn.microsoft.com/en-us/library/bb895173.aspx (accessed April 3, 2013)

13. Neto, M., G. Calvalcanti and T. Ren. 2009. Financial Time Series Prediction Using Exogenous Series and Combined Neural Networks. In Proceedings of International Joint Conference on Neural Networks June 14-19. Atlanta, Georgia

14. NeuroAI. 2013. Stock market prediction. http://www.learnartificialneuralnetworks.com/stockmarketprediction.html (accessed April 3, 2013)

15. Ortiz-Rodriguez, J. M., M. R. Martinez-Blanco, J. M. Varamontes and H. R. Vega-Carrillo. 2013. Robust Design of Artificial Neural Networks Methodology in Neutron Spectrometry. (K. Prof. Suzuki, Ed.) *Artificial Neural Networks - Architectures and Applications*

16. Pan, H., C. Tilakaratne and J. Yearwood. 2005. Predicting Australian Stock Market Index Using Neural Networks Exploiting Dynamical Swings and Intermarket Influences. *Journal of Research and Practice in Information Technology* 37(1)

17. Synergy Systems Ltd. 2013. MyStocks – Price list and trading summary for Friday, December 28, 2012. http://live.mystocks.co.ke/price_list/20121228 (accessed April 8, 2013)

18. Wong, C., and M. Versace. 2012. CARTMAP: A Neural Network Method for Automated Feature Selection in Financial Time Series Forecasting. *Neural Computing and Applications* 21(5):969-977

19. Yahoo Finance. 2013. Yahoo Finance. http://finance.yahoo.com/q/hp?s=ALU (accessed April 8, 2013)

20. Zarandi, M. H., E. Hadavandi and I. B. Turksen. 2012. A Hybrid Fuzzy Intelligent Agent-BasedSystem for Stock Price Prediction. *International Journal of Intelligent Systems* 27(11):947-969

21. Zhang, J., H. S. Chung and W. Lo. 2008. Chaotic Time Series Prediction Using a Neuro-Fuzzy System with Time-Delay Coordinates. *IEEE Transactions on Knowledge and Data Engineering* 20(7)